Reductions in Depth-integrated Picophytoplanktonic Photosynthesis Due to Inhibition by Ultraviolet and Photosynthetically Available Radiation: Global Predictions for *Prochlorococcus* and *Synechococcus*

Patrick J. Neale[1*] and Brian C. Thomas[2]

[1]Smithsonian Environmental Research Center, Edgewater, MD

[2]Washburn University, Department of Physics and Astronomy, Topeka, KS

[*]Corresponding author: P.O. Box 28, Edgewater MD, 21037, USA

nealep@si.edu

Tel: 443-482-2285

Fax: 443-482-2380

Abstract:

Phytoplankton photosynthesis in most natural waters is often inhibited by ultraviolet (UV) and intense photosynthetically available radiation (PAR) but the effects on ocean productivity have received little consideration aside from polar areas subject to periodic enhanced UV-B due to depletion of stratospheric ozone. A more comprehensive assessment is important for understanding the contribution of phytoplankton production to the global carbon budget, present and future. Here we consider responses in the temperate and tropical mid-ocean regions typically dominated by picophytoplankton including the prokaryotic lineages, *Prochlorococcus* and *Synechococcus*. Spectral models of photosynthetic response for each lineage were constructed using model strains cultured at different growth irradiances and temperatures. In the model, inhibition becomes more severe once exposure exceeds a threshold ($E_{max}$) related to repair capacity. Model parameters are presented for *Prochlorococcus* adding to those previously presented for *Synechococcus*. The models were applied to the estimation of mid-day, water-column photosynthesis based on an atmospheric model of spectral incident radiation, satellite-derived spectral water transparency and sea surface temperature. Based on a global survey of inhibitory exposure severity, a full latitude section of the mid-Pacific and near-equatorial region of the east-Pacific were identified as representative regions for prediction of responses over the entire water column. Comparing predictions integrated over the water column including versus excluding inhibition, production was 7-28% lower due to inhibition depending on strain and site conditions. Inhibition was consistently greater for *Prochlorococcus* compared to two strains of *Synechococcus*. Considering only the surface mixed layer, production was inhibited 7-73%. Weighted by lineage abundance and daily PAR exposure, inhibition of full water column production averages around 20% for the modeled region of the Pacific with two-thirds of the inhibition due to UV. The results suggest more consideration is needed on inhibition effects, especially due to UV, which are largely excluded from present models of global phytoplankton production.

**Introduction**

Phytoplankton photosynthesis in the upper layer of the ocean accounts for most marine productivity and constitutes about half of the plant productivity on Earth (Behrenfeld *et al.*, 2006). Photosynthetic performance is dependent on the penetration of photosynthetically available radiation (PAR, 400-700 nm) into the ocean water column, but is negatively affected by exposure to excessive PAR and ultraviolet radiation (UVR) in near surface waters. Ambient PAR and UVR are sufficiently intense to cause near-surface photoinhibition in most natural waters at least episodically (Harrison & Smith, 2009, Villafañe *et al.*, 2003). Studies of the spectral dependence of UV effects show that most inhibition is caused by UVA (315-400 nm) with UVB (280-315 nm, lower limit at Earth's surface, 290 nm) making a secondary contribution (Harrison & Smith, 2009, Neale, 2000). Ozone depletion increases incident UVB and has been estimated to possibly enhance inhibition by 5-12% but studies making such estimates are mostly confined to the regions of the Southern Ocean and coastal Antarctic waters affected by the seasonal appearance of the "ozone hole" (Smith & Cullen, 1995).

To assess the effect of inhibition on aquatic productivity it is necessary to take into account transmission of light into the water column, the response of photosynthesis to different wavelengths of UV and PAR and integrate effects over depth. This can be accomplished using a combination of biological weighting functions for UV-PAR inhibition of photosynthesis and photosynthesis-irradiance curves in the PAR, the so-called BWF/P-E model (Cullen *et al.*, 1992, Neale, 2000). Initial efforts to estimate BWFs were directed towards polar assemblages and species, later studies were made of temperate coastal and freshwater phytoplankton (Neale & Kieber, 2000). However little is known about the sensitivity and spectral dependence of inhibition for phytoplankton in non-polar mid-ocean waters, in which the dominant forms are picophytoplankton, particularly the prokaryotic lineages

*Synechococcus* and *Prochlorococcus*. Widely distributed in marine waters, these two lineages are estimated to account for 25% of global ocean primary productivity (Flombaum *et al.*, 2013). Neale et al. (2014) developed a new form of the BWF/P-E model ($E_{max}$ model) that accounts for the distinctive features of picophytoplanktonic photosynthetic response to PAR and UV radiation which was applied to *Synechococcus*.

Variations in both the extent of UV exposure and sensitivity of biota to UV effects may occur as the marine environment is affected by climate change (Hader *et al.*, 2015). The BWF/P-E model provides a way to assess how climate change may affect the impact of inhibition on productivity, but assessments of global change effects at regional or global scales have been largely limited to the Southern Ocean (Arrigo, 1994, Moreau *et al.*, 2015) or particular sites (e.g. estuaries or lakes) where the BWFs of the assemblage had been experimentally determined (Hiriart & Smith, 2005, Neale, 2001). Much less is known about other parts of the open ocean. The most popular model for assessing global phytoplankton production, the VGPM model of Behrenfeld and Falkowski (Behrenfeld & Falkowski, 1997) implicitly includes PAR inhibition, but not UV inhibition. It is based on the MARMAP database of simulated in situ incubations conducted in polycarbonate bottles which exclude almost all UV irradiance (O'Reilly & Thomas, 1983). To facilitate calculations of integrated production including UV/PAR inhibition on a global basis, Cullen *et al.* (2012) defined efficient numerical approximations to BWF/P-E models in which inhibition follows the irradiance-dependent "E" exposure-response relationship. This relationship is based on an equilibrium reached between the rate of UV/PAR dependent damage and repair the latter being proportional to the number of damaged targets (Neale, 2000). Using this approach, Moreau *et al.* (2015) assessed the effect of climate change on the inhibition of productivity in the coastal waters of the Antarctic peninsula where the applicability of the E model was demonstrated by Fritz *et al.* (2008). The Cullen *et al.* (2012) approximations do not consider the alternative $E_{max}$ model in which repair is limited to a maximum rate above an exposure threshold (Neale *et al.*, 2014).

In this report we present estimates of the extent to which productivity of picophytoplankton are inhibited by near surface irradiance conditions occurring in the temperate and tropical ocean. Assessments have been performed using environmental conditions applicable to broad areas of the Earth's oceans and using BWFs experimentally determined using model strains of *Synechococcus* (Neale *et al.*, 2014) and *Prochlorococcus* (results presented herein). The photosynthesis model also accounts for changes in spectral PAR at depth based on pigment absorption spectra. Effects have been evaluated over selected latitude and longitude ranges in the Pacific Ocean where we estimated spectral water transparency from satellite observations of ocean color.

**Materials and Methods:**

*BWF determination*

Productivity is modeled based on the photosynthetic responses of several strains of picophytoplankton grown in laboratory culture. Detailed descriptions of methods for culturing, photosynthesis measurements and estimation of the BWFs have been presented by Neale *et al.* (2014). They also presented the results for *Synechococcus* strains CCMP1334 and CCMP2370 (synonymous with WH7803 and WH8102). Here, we present the results for *Prochlorococcus* (strain MED4 ) which used identical methods. Strain MED4 was isolated from surface waters of the Mediterranean and is a "high-light" adapted strain representative of genotypes occurring in surface waters (Moore & Chisholm, 1999). Briefly, strain MED4 was obtained from the National Collection of Marine Algae and Bacteria (NCMA strain CCMP1986) and grown using Pro99 media (Moore *et al.*, 2007) made up in 0.2 μm filtered Sargasso Sea water at two temperatures, 20° C and 26° C, and two PAR levels, ML (18 W m$^{-2}$, ca. 77 μmol photons m$^{-2}$ s$^{-1}$) and HL (41 W m$^{-2}$, ca. 174 μmol photons m$^{-2}$ s$^{-1}$), with a 14h:10h light:dark cycle and semi-continuous dilution. Photosynthesis per unit chlorophyll (Chl) a ($P^B$) was measured as total $^{14}$C assimilation (acid-stable) during 1 hour incubation in a custom designed spectral incubator ("photoinhibitron") as described by Neale et al. (2014). Similar to the previous experiments with *Synechococcus*, before the measurement of photosynthesis the culture suspension was pretreated by one hour exposure to moderately high PAR (400 μmol m$^{-2}$ s$^{-1}$) and UV from a xenon source filtered to exclude wavelengths < 350 nm [details in (Neale *et al.*, 2014)]. The motivation behind this treatment, together with relatively high growth irradiances, was to induce and activate inhibition defense mechanisms as much as possible so that BWF sensitivity estimates would be conservative. Culture aliquots (1 mL) were then exposed to 10 irradiance levels each of 12 spectral treatments defined using long-pass filter combinations with 50% Transmission wavelengths varying from 291 to 408 nm. Spectral irradiance (mW m$^{-2}$ nm$^{-1}$) for each well in the photoinhibitron was measured with a custom built fiber-optic spectroradiometer as described by Neale & Fritz (2001). The results were fit to the BWF/P-E$_{max}$ model defined by the equations:

$$P^B = P_s^B \cdot \left(1 - e^{-E_{PAR}/E_s}\right) \cdot ERC(E_{inh}^*) \quad (1)$$

$$ERC(E_{inh}^*) = \begin{cases} 1/(1+E_{inh}^*) & E_{inh}^* \leq E_{max}^* \\ 1/cE_{inh}^* & E_{inh}^* > E_{max}^* \end{cases}$$

$$c = \frac{1 + E_{max}^*}{E_{max}^*} \quad (2)$$

Where $P^B_s$ (mg C mg Chl$^{-1}$ h$^{-1}$) is the maximum rate of photosynthesis normalized to chlorophyll (Chl, mg Chl m$^{-3}$) biomass, $E_s$ (W m$^{-2}$) is the saturation parameter for PAR irradiance ($E_{PAR}$) and ERC, the exposure response curve for inhibition of photosynthesis, is a function of total inhibitory irradiance, $E^*_{inh}$ defined by

$$E_{inh}^* = \sum_{265\,nm}^{400\,nm} \varepsilon(\lambda) \cdot E(\lambda) \cdot \Delta\lambda + \varepsilon_{PAR} \cdot E_{PAR} \quad (3)$$

Where the set of $\varepsilon(\lambda)$ is the Biological Weighting Function (BWF) and $E(\lambda)$ is spectral UV irradiance (see Supplementary Table 1 for a full list of symbols and units). A single parameter, $\varepsilon_{PAR}$ (W m$^{-2}$)$^{-1}$, accounts for inhibition by PAR. The threshold parameter $E^*_{max}$ defines the transition between the exposures for which repair increases with damage ($E^*_{inh} \leq E^*_{max}$) to higher exposures ($E^*_{inh} > E^*_{max}$) for which repair is constant (i.e. operating at some maximum rate). Operationally, the relative effect of inhibition follows a rectangular hyperbolic response at low exposure transitioning to an inverse response above $E^*_{max}$, for more details see Neale *et al.* (2014).

These experiments included some treatments with $E(\lambda)$ with $\lambda < 290$ nm in order to account for effects of severe ozone depletion due to astrophysical ionizing radiation, the results of which are discussed by Neale & Thomas (2016). Trial fits with and without these treatments showed that their inclusion does not affect the BWF at $\lambda > 290$ nm (Neale *et al.*, 2014). Experiments were repeated at least three times with independently grown cultures for each set of conditions. Cell enumeration was performed with a Multisizer 4 particle sizer/counter (Beckman/Coulter) using a 20 µm aperture (minimum resolution 0.4 µm). Samples were diluted as necessary with 0.2 µm filtered seawater before counting.

*Productivity Profiles*

Productivity calculations were driven by estimates of the depth profile of spectral irradiance (290-750 nm). These estimates were produced in a three-step process previously described by Thomas *et al.* (2015) and Neale & Thomas (2016). First, an atmospheric chemistry and radiative transfer model was run to generate surface incident irradiance at midday (Thomas *et al.*, 2015). Model results were used for atmospheric chemistry instead of actual observations so that the effects of UV could be compared to other model runs in which atmospheric chemistry was perturbed by astrophysical ionizing radiation. The results of these comparisons are presented in another report, (Neale & Thomas, 2016). Nevertheless, we used the atmospheric chemistry model results for the "normal" atmosphere which are representative of present day conditions for the radiative transfer model (Thomas *et al.*, 2015). The resulting direct and diffuse irradiance components were transferred through the ocean surface following the procedure outlined in Arrigo et al. (Arrigo *et al.*, 2003). Spectral irradiance at depth was calculated using diffuse attenuation coefficients ($K_d(\lambda)$) estimated using the SeaWiFS remotely-sensed reflectance climatology (2010 reprocessing). Visible range $K_d$'s (412-700 nm, 1 nm resolution) were estimated using the Quasi-Analytical Algorithm (QAA) v6 (Lee *et al.*, 2013). A second set of $K_d(\lambda)$ at $\lambda = 320, 340, 380, 412, 443$, and $490$ nm were estimated using the SeaUV algorithm (Fichot *et al.*, 2008). The two sets were merged by adjusting the SeaUV $K_d$'s to agree with QAA v6 output at 412 nm, with the magnitude of the offset proportionally reduced with decreasing wavelengths to 0 at 320 nm [cf. (Lee *et al.*, 2013)], and interpolation/extrapolation based on log transformed values to 290-412 nm (1 nm resolution). The model then calculated photosynthesis rates for each phytoplankton species using the BWF/P-E model as described, resulting in photosynthesis rates at a range of depths in the water column, as well as a depth-integrated value. The calculation computed potential photosynthesis rates (no inhibition, ERC=1), and with inhibition by both UV and PAR. The $E_{PAR}$ used to estimate in situ rates was adjusted to account for differences in photosynthetically utilizable radiation (PUR) between in situ irradiance and the photoinhibitron as previously described (Neale *et al.*, 2014). The adjusted irradiance is denoted $E'_{PAR}$.

Since atmospheric chemistry calculation was only 2-D (no longitude, Thomas *et al.* 2015) and the model is overall numerically intensive, the full simulation covered only selected regions of the ocean. These corresponded to the longitude range between 160°W and 140°W, and 10 degree latitude bands centered at 85°S to 85°N, which is an area of the mid-Pacific Basin near the longitude of the Hawaiian Islands; and a longitude range between 110°W and 90°W, and bands centered between 15°S and 15°N in latitude, a region near the longitude of the Galapagos Islands. The northern part of the latter region includes an area of upwelling known as the Costa Rica Dome. These regions were chosen to be representative of the overall variation in exposure variation and phytoplankton biomass in Earth's ocean based on transparency depths (discussed in next section). All other longitude-dependent data were then processed using the same regions.

*Transparency depths*

We conducted a global survey of near-surface irradiance conditions in the ocean using metrics for inhibiting and photosynthetically utilizable irradiance, $T_{PIR}$ and $T_{PUR}$, respectively. These were computed following Lehmann *et al.* (2004), except that irradiance was in energy (W m$^{-2}$) units. These metrics gauge the impact of solar radiation on the profile of phytoplankton photosynthesis taking into account both incident radiation and water column transparency. Transparency to inhibiting irradiance ($T_{PIR}$) is defined as:

$$T_{PIR} = \sum_{290nm}^{700nm} \frac{1}{K_d(\lambda)} \varepsilon(\lambda) E(0^-,\lambda) \Delta\lambda \qquad (3)$$

with $\varepsilon(\lambda)$ in the PAR region set to a constant value $\varepsilon_{PAR}/300$. The selection of BWFs for the calculation was based on surface layer conditions as discussed below. The transparency to photosynthetically utilizable radiation ($T_{PUR}$) is defined as

$$T_{PUR} = \sum_{400nm}^{700nm} \frac{1}{K_d(\lambda)} \frac{a_p(\lambda)}{\bar{a}_p} \frac{E(0^-,\lambda)}{E_{PAR}(0^-)} \Delta\lambda \qquad (4)$$

where $a_p(\lambda)$, (m$^2$ mg Chl$^{-1}$), is the chlorophyll-specific spectral absorption coefficient for each strain used in the simulation (Neale *et al.*, 2014). $T_{PIR}/T_{PUR}$ is the average inhibition weight over the first attenuation length for photosynthetically utilizable radiation, in other words, a metric for how much inhibiting exposure phytoplankton in the euphotic zone receive. Previous analyses have shown that $T_{PIR}/T_{PUR}$ is highly correlated with percent inhibition of depth integrated, instantaneous water column photosynthesis estimated by a full numerical integration (Lehmann *et al.*, 2004, Neale, 2001). Our survey was based on incident irradiance (a function of time and latitude only) and the global $K_d(\lambda)$ climatology of Fichot and Miller (2010).

*Selection of BWF/P-E parameters*

Parameters used in the model calculations were chosen by reference to estimated water column temperature and average irradiance for each date and location. Sea surface temperatures were retrieved from the IRI/LDEO Climate Data Library (http://iridl.ldeo.columbia.edu).  Specifically, we used the Reynolds and Smith (Reynolds & Smith, 1995) global monthly climatologies.  We again averaged over 10° latitude bands, and the longitude ranges described above and considered the surface temperature to apply to whole mixed layer.

Light saturated rate of photosynthesis ($P^B_s$) and light saturation parameter ($E_s$) are well known to vary with temperature according to an Arrhenius equation ($Q_{10}$) type response (Geider & MacIntyre, 2002). Based on this assumption, temperature ($T$) functions were estimated for $P^B_s$ and $E_s$ using an equation of the form $m_1 \exp(m_2 (T-20))$.  $P^B_s$ and $E_s$ were not significantly different between the two growth irradiances (Neale et al. 2014 and results below), so all growth data for each temperature were pooled in estimating the equations (n=12-14).  As expected, in most cases $m_2$ was not significantly different from 0.0693 corresponding to a $Q_{10}$ of 2, (results not shown, cf. (Eppley, 1972)). These functions were applied for temperatures below 26°C, above that, parameters for 26°C growth conditions were used. To obtain weighted inhibiting irradiance ($E^*_{inh}$) appropriate to the sea surface temperature between 20° and 26°C, average $E^*_{inh}$ was calculated for each temperature and linearly interpolated.  The average $E^*_{inh}$ for 20° was used for lower temperature and that for 26° at higher temperatures.

To choose between BWFs obtained for the two growth irradiances, we first computed the average PAR irradiance in the upper ocean, defined by the mixed layer depth.  Monthly climatologies of mixed layer depth based on density profiles were retrieved from the Naval Research Laboratory website (http://www7320.nrlssc.navy.mil/nmld/nmld.html).  This global data was averaged over 10° latitude bands, and the longitude regions described above.  The average upper ocean PAR was then compared to 140 µmol photons m$^{-2}$ s$^{-1}$ which is an intermediate level between ML and HL irradiance BWF versions; for values less than, the ML BWF version was used, while for values greater, the HL BWF version was used.

**Results:**

*BWFs for Inhibition of Photosynthesis in* Prochlorococcus

Maximum rate of uninhibited photosynthesis ($P^B_s$) and saturation irradiance parameter ($E_s$) were higher for *Prochlorococcus* cultures grown at higher temperature (Table 1). As previously observed for *Synechococcus*, there were no consistent differences in these parameters for the two different irradiance conditions (*t*-test, p>0.05). The parameter for inhibition by $E_{PAR}$, $\varepsilon_{PAR}$, also tended to be lower for cultures grown at higher irradiance, but not significantly so. Culture growth rates were in the range previously reported for MED4 (Fu *et al.*, 2007, Kulk *et al.*, 2012, Moore & Chisholm, 1999). Average BWFs±SE (n ≥ 3) for *Prochlorococcus* (MED4) are shown in Fig. 1.

The BWFs varied between growth conditions in a similar way as previously observed for *Synechococcus* (WH8102). In general, the average BWF for ML cultures showed greater sensitivity (larger weights) than those for HL cultures, but the differences were close to the standard error of the determinations (includes statistical uncertainty of estimates and replicate uncertainty between cultures). The difference between HL and ML was larger at the growth temperature of 26°C, but HL cultures also had a significantly lower $E^*_{max}$ (*t*-test, p=0.02). The implication of this difference in $E^*_{max}$ is best understood in relation to a specific spectral exposure. Thus, we estimated how much solar exposure would be required for weighted irradiance to reach $E^*_{max}$ for each culture condition (Table 1, last column). For HL cultures, $E^*_{max}$ is reached at approximately the same solar exposure for both growth temperatures, i.e. within the uncertainty of the estimate. This situation arises because the $E^*_{max}$ threshold is relative to weighted exposure ($E^*_{inh}$, see Eq. 2) which is, in turn, a function of the BWF. Since both the BWF and $E^*_{max}$ are lower at the higher temperature the threshold is approximately the same in terms of absolute exposure (Table 1). Effectively, this means that below the exposure threshold of approximately 20% of full sun, the 26°C cultures are less sensitive than 20°C cultures but above the threshold sensitivity is higher and about the same for both growth temperatures. While growth conditions affected photosynthetic and inhibition sensitivity responses similarly in *Synechococcus* and *Prochlorococcus*, overall *Prochlorococcus* had higher BWF weights than either *Synechococcus* strain (Fig. 1d). Notably, this is for a "high-light" adapted strain of *Procholoroccocus* (MED4) and low-light adapted strains that are more abundant in the thermocline and below are probably more sensitive [cf. (Moore & Chisholm, 1999, Six *et al.*, 2007)].

Predicted profiles of midday photosynthesis at 25°N in the mid-Pacific zone (140 to 160°W) provide a good example of the joint effect of the variation in all the BWF/P-$E_{max}$ parameters for *Prochlorococcus* under typical in situ conditions (Fig. 2). Climatological monthly averages were calculated using model irradiance and observed temperatures as described in *Materials and Methods*. Temperature varied between 21.8°C in February to 25.4°C in September. The calculations do not take into account changes in photosynthetic response below the thermocline, so deep rates are mainly presented for comparative purposes. Figure 2 compares predicted profiles of $P^B(z)$ (biomass normalized photosynthesis as a function of depth) and $P^B_{pot}(z)$ which shows the potential rate in the absence of inhibition (i.e. ERC=1). The latter set of profiles show the positive effect of higher temperature and more incident irradiance in the summer (June-Sept) vs winter period. The light-saturated rate of photosynthesis ($P^B_s$) is higher and noon-time photosynthesis extends beyond 180 m in the summer while there is negligible activity at this depth in the winter. At the same time, the $P^B(z)$ profiles show that inhibition effects are also much stronger and affect more of the profile in the summer. Two indicators of this are (1) the depth at which exposure reaches $E^*_{max}$ (indicated by the upper horizontal line in Fig. 2) and (2) the maximum depth of inhibition (i.e. at which $P^B(z)$ and $P^B_{pot}(z)$ profiles diverge, horizontal lower line Fig. 2). From winter to summer, the $E^*_{max}$ threshold depth increases from 10 m to more than 20 m and depth at which inhibition exceeds 0.1 mg C mg Chl$^{-1}$ h$^{-1}$ increases from around 50 m to 80 m. While clearly stronger in the summer, inhibition moderates, but doesn't completely negate, the enhancement in depth integrated production ($P^B_Z$) due to higher temperature and more light.

Production per unit Chl, integrated to 0.1% depth for $E'_{PAR}$ is higher in the summer than winter (e.g 193 vs 127 g C g Chl m$^{-2}$ h$^{-1}$ for September vs January, respectively).

*Global Survey of Ocean Transparency to inhibitory UV+PAR*

Estimated annual average $T_{PIR}/T_{PUR}$ ratio, a metric for inhibiting exposure in the upper euphotic zone, is shown using global color contour maps for all three modeled picophytoplankton strains in Figure 3. For convenience, we will use the abbreviation "*Pro*" for *Prochlorococcus* MED4, "*Syn1*" and "*Syn2*" for *Synechococcus* WH8102 and WH7803, respectively. *Pro* has the highest sensitivity to UV and highest average $T_{PIR}/T_{PUR}$ (maximum 3). The ratio is lower by about half for *Syn1* (max 1.75) and *Syn2* (max 1.5). Receiving the highest irradiance and having the highest UV transparency, the subtropical gyres of the oceans show the regionally greatest $T_{PIR}/T_{PUR}$. The equatorial region is lower than the subtropics, especially in the Eastern Pacific due to lower transparency in regions with equatorial upwelling. The average ratio decreases by about a factor of 3 between the tropics and the high latitudes (this analysis does not include the Arctic and Antarctic). The contours on the $T_{PIR}/T_{PUR}$ ratio maps show that the overall spatial pattern in inhibitory exposure to the euphotic zone is relatively similar between strains, although different in magnitude. Based on these results we chose two exposure regions as representative of the full range of inhibition stress within the global ocean. These correspond to the longitude range between 160°W and 140°W, at all latitudes (mid-Pacific region); and a longitude range between 110°W and 90°W, between 15°S and 15°N in latitude where there is substantial equatorial upwelling (E-Pacific region). The annual zonal average $T_{PIR}/T_{PUR}$ in these regions fall within the range of global averages, with region 2 providing the best match to global averages in the equatorial region (Neale & Thomas 2016). These regions were then used to examine the seasonal and latitudinal dependence of inhibition effects for each modeled strain.

*Integrated picophytoplankton productivity*

To examine the overall effect of UV+PAR inhibition on picophytoplankton production, we calculated the ratio of production using the full BWF/P-E model ($P^B_Z$) to production in the absence of inhibition ($P^B_{Z\text{-POT}}$), in each case integrated to the depth of 0.1% $E'_{PAR}$. These integrals correspond to the left and right members of the profile pairs shown in Fig. 2. A 24-month sequence based on climatological data is shown to visualize the unbroken seasonal sequence in each hemisphere. Depending on strain, time of year and latitude the $P^B_Z$: $P^B_{Z\text{-POT}}$ ratio varied between 0.93 and 0.72, i.e. full depth profile productivity was inhibited from 7 to 28% (Fig. 5). Effects by strain ranked similarly as the global map of $T_{PIR}/T_{PUR}$, *Pro* productivity is most affected by inhibition, up to 28%, while *Syn2* showed the least effects.

There were also differences between strains in the space-time distribution of inhibition. Within each hemisphere in the mid-Pacific (Fig. 4a-c), productivity by *Pro* and *Syn2* was most inhibited in the two months before the summer solstice (i.e. April and May in the NH), whereas for *Syn1* inhibition was most pronounced around the solstice (June and July). Over the latitude range, peak inhibition was at 30°S for *Pro* and *Syn2*, but around 20°S for *Syn1*. Maxima in inhibition are also observed around the same latitudes in the NH, but at lower magnitude. Although $T_{PIR}/T_{PUR}$ is not particularly greater in the SH vs NH (Fig. 3), the magnitudes of $T_{PIR}$ are about 50% higher at the SH maximum (Neale & Thomas, 2016). Such high transparencies accentuate the inhibition effect since a greater proportion of exposure is above $E^*_{max}$. The pattern of inhibition in the SH and NH also differs quite markedly for *Syn1*, in the SH there is a greater contrast in inhibition effect between low and high latitudes. This contrast is not as sharp for *Syn2* and *Pro*.

In the E. Pacific region, the general pattern of sensitivity and seasonal variations of response was similar to the mid-Pacific at the same latitudes with some differences in detail (Fig. 4d-f). For *Syn1* there was less effect in the E. Pacific south of the equator, but about the same north of the equator.

The opposite was true for *Pro*, with there being more effect south of the equator in the E Pacific. *Syn2* exhibited a similar pattern in both mid-Pacific and E. Pacific equatorial regions.

An alternate set of integrations was performed extending only to the depth of the upper mixed layer as opposed to the full depth range (i.e. to 0.1% $E'_{PAR}$). The full depth range estimate provides a conservative estimate of $P^B_{Z\text{-}POT}$, but, like many other models used to estimate global phytoplankton productivity, unrealistically assumes uniform physiological parameters even extending below the thermocline. Integration over only the upper mixed layer, considers the opposite extreme in which the model is only applied to portion of the water column that matches the parameter choice. In these integrations, the inhibition effect was much stronger and there was less contrast between predictions using different strain parameters (Fig. 5a-c). In all cases in the mid-Pacific, the greatest effect was at the summer solstice at 25° latitude. Again, the most intense effects are in the SH reaching as much as 73% inhibition of mixed layer production for *Pro* and more than 60% inhibition for *Syn1* and *Syn2*. This corresponds to the coincidence of high incident UV, high water transparency (indicated by high $T_{PIR}/T_{PUR}$ ratio) and (relatively) shallow mixed layer depths at this time and location. Effects at the SH summer solstice decrease considerably both to high and low latitude side of this inhibition peak. On the high latitude side this is mainly due to lower transparency and deeper mixed layer depth at higher latitudes. The decrease at high latitude is particularly steep for *Syn1*. The decrease on the low latitude side is seemingly counterintuitive since incident irradiance is near or at the maximum in this location. But high incident irradiance is more than counterbalanced by deeper mixed layer depths and lower transparency near the equator compared to the tropics and subtropics (Pennington *et al.*, 2006). However, equatorial mixed layer depths shoal further to the East (Pennington *et al.*, 2006), as a result effects on mixed layer inhibition is higher in the E. Pacific compared to the same latitude in the mid Pacific (Fig. 5d-f).

In order to estimate the overall effect of UV and PAR inhibition on a hemispheric basis we averaged production weighting by the biomass distribution of the *Procholorococcus* and *Synechococcus* lineages and a daily productivity multiplier for the times and sites used for the mid-day productivity calculation. We estimated surface abundance of each lineage using the equations of Flombaum *et al.* (2013). These authors parameterized *Procholorococcus* and *Synechococcus* (sensu lato) abundance in the ocean as a function of water temperature and daily PAR. Expected abundances for our times and locations were estimated using the Flombaum *et al.* (2013) parameterization given the monthly temperature climatology and the SeaWIFs daily PAR climatology (2010 reprocessing). The daily productivity multiplier was estimated as the ratio (in mol photons m$^{-2}$) of SeaWifs total daily PAR to the 1 hour midday PAR, the latter from the incident PAR used in the productivity calculation. A weighted average monthly production ($<P>_Z$) was calculated over the hemisphere by weighting $P^B_Z$ at each latitude by the corresponding lineage abundance and the productivity multiplier. Since irradiance data is seasonally absent from the SeaWiFS dataset in the high latitude winter, only bands up to 45° were included. The abundance of picophytoplankton outside of this band is typically low (Flombaum *et al.* 2013). The calculation combined the two near-equatorial productivity estimates in mid-Pacific and east Pacific by further weighting each by 0.5 so that each region contributed equally to the overall average at those latitudes. A similar calculation was performed using $P^B_{Z\text{-POT}}$, thus obtaining $<P>_{\text{pot-Z}}$. The average effect of UV+PAR inhibition over the hemisphere was then taken as 1- $<P>_Z/<P>_{\text{pot-Z}}$. Relative inhibition by this measure is in the range of 15-45%, depending on the hemisphere, BWF and integration depth (Fig. 6). In all cases, the severity of inhibition by strain is in the order *Pro*> *Syn*1 > *Syn*2. In the Northern Hemisphere (NH), inhibition of production integrated to 0.1% LD averages around 20% through most of the year. This reflects the weighting by lineage abundances. For *Synechococcus*, the peak abundance is in the sub-tropics and seasonally tracks the position of the 80% $P^B_Z$: $P^B_{Z\text{-POT}}$ ratio of *Syn1* and *Syn2* (cf. Flombaum *et al.* 2013). For *Prochlorococcus* the peak abundance is closer to the equator, but again the average *Pro* inhibition in the NH tropics (0°-25°) does not vary much with time (Fig. 4c &f). In the Southern Hemisphere (SH), inhibition of 0.1% LD production shows more seasonality. Because of the strong summertime inhibition around 25°S (Fig. 4), average inhibition is several percent higher in the SH summer compared to the NH summer (Fig. 6b & d). As would be expected, average inhibition is much greater for production integrated over the MLD and shows strong seasonality. Shallow mixed layer depths combined with strong inhibitory irradiance cause inhibition to peak during the summer solstice in both hemispheres. Again, the highest average inhibition is for *Pro* in the late summer SH.

A separate set of integrations was performed which only included inhibition by PAR ($P^B_{Z\text{-}PAR}$). These were compared to the integrations with full spectral (UV+PAR) inhibition in order to estimate the relative contribution of PAR and UV to inhibition. Specifically, we compared the reduction of $P^B_{Z\text{-}PAR}$ relative to $P^B_{Z\text{-}POT}$ as a proportion of $P^B_{Z\text{-}POT} - P^B_Z$. The proportion of inhibition due to PAR so calculated and weighted averaged as described above, was fairly consistent between strains. The proportion was 33±3% (mean±SD) for integrations to 0.1% $E'_{PAR}$ and 31±2% for integrations to the ML depth. In other words, UV was responsible for about two-thirds of the predicted inhibition.

**Discussion:**

Biological weighting functions have been estimated for representative strains of the most common prokaryotic picophytoplankton in the open ocean, *Synecchococcus* and *Prochlorococcus*, and show that these lineages are highly sensitive to inhibition of photosynthesis by UV radiation. This sensitivity translates to a considerable reduction in predicted primary productivity under conditions expected in the surface layer of temperate and tropical ocean. For conditions in the mid-Pacific, inhibition over the whole water column ranged from 7 to 28%, with hemispheric effects averaging around 20% with mild seasonality. Only considering the surface mixed layer, inhibition of integrated primary productivity ranged from 7 to 73%. Modeled estimates of hemispheric average inhibition in the mixed layer were strongly seasonal, varying from around 20% in winter to around 40% in summer (>40% for *Prochlorococcus*). On average, UV was responsible for about two-thirds of the inhibition, considered on either a full water column or mixed layer only basis. These inhibition predictions suggest that current model estimates of ocean primary productivity that do not take into account UV inhibition may overestimate production in areas where *Synecchococcus* and *Prochlorococcus* are the dominant contributors. Most models used to assess global primary production do not account for UV inhibition (Carr *et al.*, 2006). Moreover, the response of primary productivity to climate change should include how such change influences inhibition through variations in UV exposure and/or variations in sensitivity due to alterations in the marine environment.

In assessing how much confidence to place in BWF/P-E$_{max}$-based estimates of relative inhibition, we consider how reasonable are both (1) the estimation of potential productivity (in absence of inhibition) (2) the estimation of inhibited photosynthesis. Potential productivity is a function of the photosynthesis-irradiance (P-E) parameters and irradiance light fields. Our estimates of P-E parameters for *Synechococcus* (Neale *et al.*, 2014) and *Prochlorococcus* (Table 1) are consistent with previous laboratory studies on the same strains, e.g. for *Syn1* (Fu *et al.*, 2007), *Syn2* (Kana & Glibert, 1987b) and *Pro* (Fu *et al.*, 2007, Moore & Chisholm, 1999, Partensky *et al.*, 1993). These laboratory studies, as well as ours, grew cultures under nutrient replete conditions whereas nutrient limitation prevails most of the time in natural populations of picophytoplankton. Nevertheless, our estimates of maximum photosynthetic rates per cell in the absence of inhibition ($P^{cell}_s$) are comparable to rates reported for natural populations based on $^{14}$C incorporation followed by flow cytometric cell sorting. For *Pro*, $P^{cell}_s$ averages 1.52 fgC cell$^{-1}$ h$^{-1}$ (SD±0.83) over all growth conditions. Average rate for *Prochlorococcus* assemblages in the tropical N. Atlantic (20 stations) under optimum light conditions (~10 % surface irradiance, no UV, SST=23±1°C) was 1.2±0.6 fgC cell$^{-1}$ h$^{-1}$ (Jardillier *et al.*, 2010). Rates for *Prochlorococcus* assemblages in other locations are in a similar range (Li, 1994, Rii *et al.*, 2016). For *Synechococcus*, data to estimate $P^{cell}_s$ for our experiments is only available for *Syn1* grown at 20°C, which averages 6.6±4.4 fgC cell$^{-1}$ h$^{-1}$ over the two growth light levels. Jardillier *et al.* (2010) report rates of 9.5±4.3 fgC cell$^{-1}$ h$^{-1}$ for samples having a high proportion of *Synechococcus* clade III genotypes (*Syn1* is in the same clade), consistent with the somewhat higher temperature in their samples.

Our potential productivity estimates are also based on model estimates of incident and downwelling spectral irradiance which are in turn derived from the output of an atmospheric model and remotely sensed ocean color. Our calculations used more recent, and improved, approaches to estimate irradiance attenuation in the UV and visible than used in previous model estimates of UV inhibition of productivity (Arrigo *et al.*, 2003, Cullen *et al.*, 2012, Neale *et al.*, 1998). However, our estimates of potential productivity are consistent with previous work. Midday potential productivity integrated to the depth of 0.1% $E'_{PAR}$ is within 3-4% of estimates of the same quantity estimated using the Depth Integrated Model (DIM) of Cullen et al. (2012), i.e. given our reported values of $P^B_s$ and $E_s$. The relative accuracy is similar to that reported by Cullen et al. (2012) in their tests of the DIM model vs direct numerical integration. Note, for purposes of comparison, DIM model calculations used $T_{PUR}$ as described in Cullen et al. (2012), i.e. irradiance in quanta units. The agreement between these two independent methods of estimating potential productivity both validates our calculations and further reinforces the applicability of the DIM equations which heretofore have not been compared to full numerical integration beyond the original comparison conducted by Cullen et al. (2012).

Based on these comparisons we conclude that our estimates of substantial inhibition of integrated productivity are not due to inflated estimates of potential productivity, rather to the low estimated rates of production in Pacific near-surface under full spectral UV+PAR irradiance conditions. As this is a novel result, there are few other reports to which these estimates can be compared. Cullen *et al.* (2012) used the DIM to estimate inhibition of daily production ranging up to 10% for the mid-Pacific in October based on the BWF/P-E for the diatom *Phaeodactylum*. Moreau *et al.* (2015), also using the DIM, estimated that the average loss to daily production due to inhibition in the West Antarctic Peninsula region was 7±1.6%. These relative losses are considerably lower than the percentages reported here, for example in October the maxima in our estimates are 23% for *Synechococcus* and 26% for *Prochlorococcus*. Cullen *et al.* (2012) and Moreau *et al.* (2015) estimated inhibition of daily production whereas we have estimated inhibition of instantaneous mid-day productivity but Cullen *et al.* (2012) found that percent inhibition of daily production is only slightly lower than that of instantaneous midday, e.g. the upper 5th percentile of inhibition in their parameterizing data set was 10.3% vs 10% for instantaneous vs daily productivity, respectively. It is possible to use BWFs for *Synechococcus* and *Prochlorococcus* to estimate normalized inhibiting irradiance ($E^*_{PIR}$) and transmission of inhibiting irradiance ($T_{PIR}$) as needed to evaluate %inhibition using the DIM (Cullen *et al.* 2012 equation 36). However, % inhibition so estimated is more than 25% lower on average than our direct calculation based on $P_z$ vs $P_{pot-z}$. The relationship between the two estimates is linear but also has considerable scatter, with the relative underestimate (average±standard deviation) being 30±8%, 43±6% and 26±6% for *Syn1*, *Syn2* and *Pro*, respectively. This variable underestimate is probably due to the presence of a threshold in the $E_{max}$ model, exposures above which inhibition is more severe. The DIM model is based on the "E" model that lacks a threshold.

Given that inhibition relationship is linear albeit with scatter, it is likely that a streamlined computational approach similar to the DIM but appropriate for the $E_{max}$ model can be developed in the future. This will facilitate estimates of inhibition impact on primary productivity on a multiple regional up to global basis. In addition, more measurements of primary productivity under UV exposure are needed in the temperate and tropical ocean to compare with model estimates [cf. (Fuentes-Lema *et al.*, 2015)]. Previous studies on how the sensitivity of photosynthesis to UV inhibition is affected by nutrient limitation have focused on eukaryotic species which are not picophytoplankton with varying results [reviewed by Beardall et al (2014)]. Both field measurements and additional laboratory studies will be required to understand the responses of prokaryotic picophytoplankton. While *Prochlorococcus* and *Synechococcus* are important contributors to primary productivity in the open ocean, recent studies have also shown that various strains of picoeukaryotes can make a significant contribution, sometimes dominating biomass specific production (Jardillier *et al.*, 2010, Rii *et al.*, 2016). However, we presently know little about the UV sensitivity of ocean strains of picoeukaryotes.

Since BWF models account for spectral changes in irradiance and parameters can be varied for difference environmental conditions, they enable assay of inhibition effects under present day conditions as well as future conditions considering the possible effects of climate change. Given the present controls on the release of chlorofluorocarbons (Montreal Protocol), there is currently no expectation that ozone depletion on the same scale as occurs in polar regions will occur in temperate or tropical regions (UNEP, 2014). However, large changes in ozone concentration could occur due to extraterrestrial influences on earth's atmosphere such as those caused by astrophysical ionizing events (Thomas *et al.*, 2015). Neale and Thomas (2016) use the BWF/P-$E_{max}$ model to assess the possible effect of astrophysical ionizing events on marine primary productivity. Other atmospheric effects associated with climate change will also influence incident and downwelling UV in aquatic systems and thus the extent of inhibition. These include changes in cloud cover, air and water pollutants, and changes in stratification affecting the depth of the surface mixed layer (Williamson *et al.*, 2016).

Moreover climate change related changes in growth condition affect strain sensitivity to UV. These include changes related to photoadaptation, nutrient supply and ocean acidification (Beardall *et al.*, 2014). As future work defines how BWF/P-E responses of picophytoplankton change in relation to these perturbations, the modeling framework presented here will enable assessing the effects on ocean productivity.

**Acknowledgements**

This research was supported by NASA grant NNX09AM85G to Brian Thomas, Patrick J. Neale and Adrian Melott. The authors thank Alicia Pritchard and Ryan Ihnacik for laboratory assistance and Pedro Flombaum for sharing his Matlab code.

Figure 1 – Average (±standard error) of biological weighting functions for *Prochlorococcus* (MED4) grown in either "medium" irradiance (ML, 18 W m$^{-2}$/77 µmol m$^{-2}$ s$^{-1}$ PAR) or "high" irradiance (HL, 41 W m$^{-2}$/174 µmol m$^{-2}$ s$^{-1}$ PAR), and growth temperatures of 20° or 26° C.

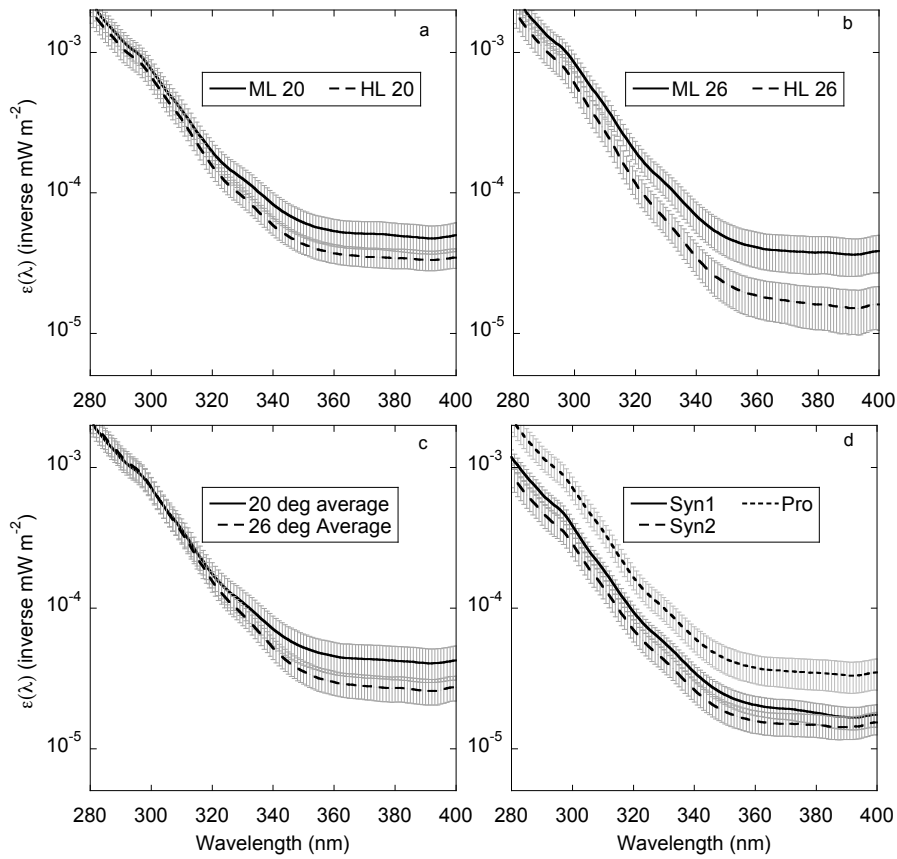

Figure 2  Predicted monthly average $P^B$ profiles for noon at 25°N, mid-Pacific longitudes (140° to 160°W) based on the BWF/P-$E_{max}$ model using parameters for *Prochlorococcus*.  Two profiles are shown for each month, a profile including inhibition by UV+PAR ($P^B(z)$ - solid line), and a profile of potential photosynthesis excluding inhibition ($P^B_{pot}(z)$ - dashed line).  Vertical grid lines are spaced at intervals of 2 mg C mg Chl$^{-1}$ h$^{-1}$.  Fine dashed lines trace the depth at which exposure exceeds $E^*_{max}$ (upper line) and the maximum depth of inhibition, i.e. the divergence between $P^B(z)$ and $P^B_{pot}(z)$ profiles > 0.1 mg C mg Chl$^{-1}$ h$^{-1}$ (lower line)

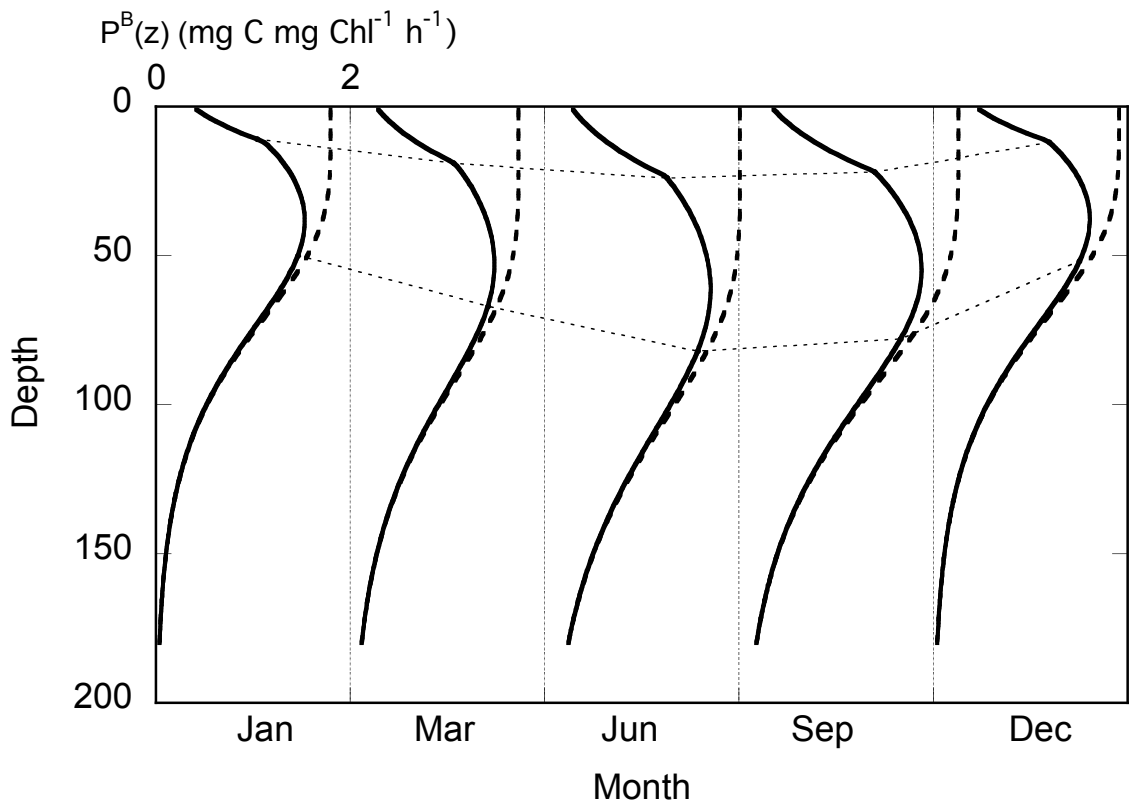

Figure 3- Global Map of $T_{PIR}/T_{PUR}$ ratio indicating the average inhibition weight over the upper attenuation length for $E'_{PAR}$ based on the BWFs and cellular absorbance of *Prochlorococcus* (Pro), and *Synechococcus* strains WH8102 (Syn1) and WH7803 (Syn2). Rectangles indicate areas selected for running the full model in the mid-Pacific near the longitude of the Hawaiian Islands and equatorial East Pacific near the longitude of the Galapagos Islands.

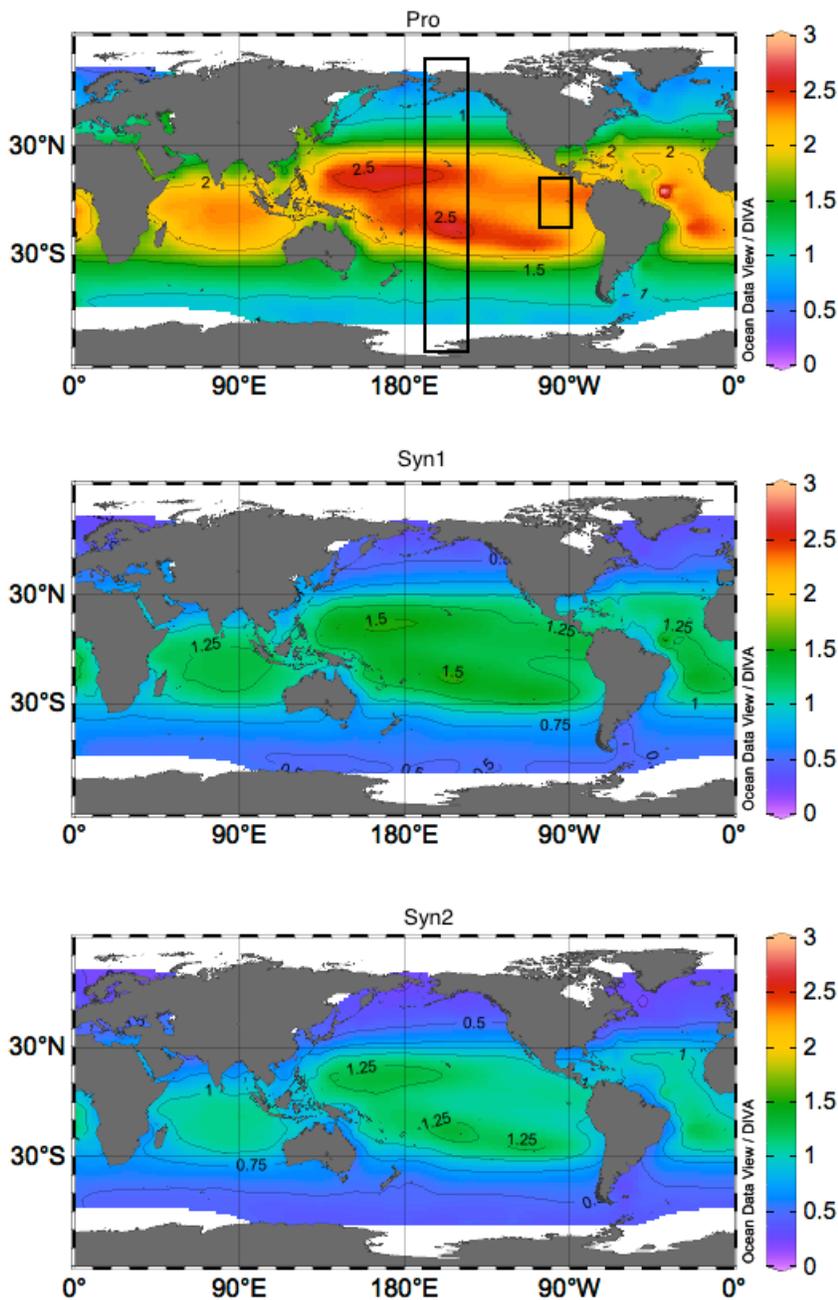

Figure 4 - Multipanel time-latitude contour plots showing $P^B_Z : P^B_{Z\text{-POT}}$ ratio for mid-day depth integrated to the 0.01% light depth for $E'_{PAR}$ for a 24 month period staring in June. Upper panels show zonal averages for the mid-Pacific (longitude of Hawaii) and lower panels for the east Pacific, near the equator (longitude of Galapagos, cf. Fig. 3).

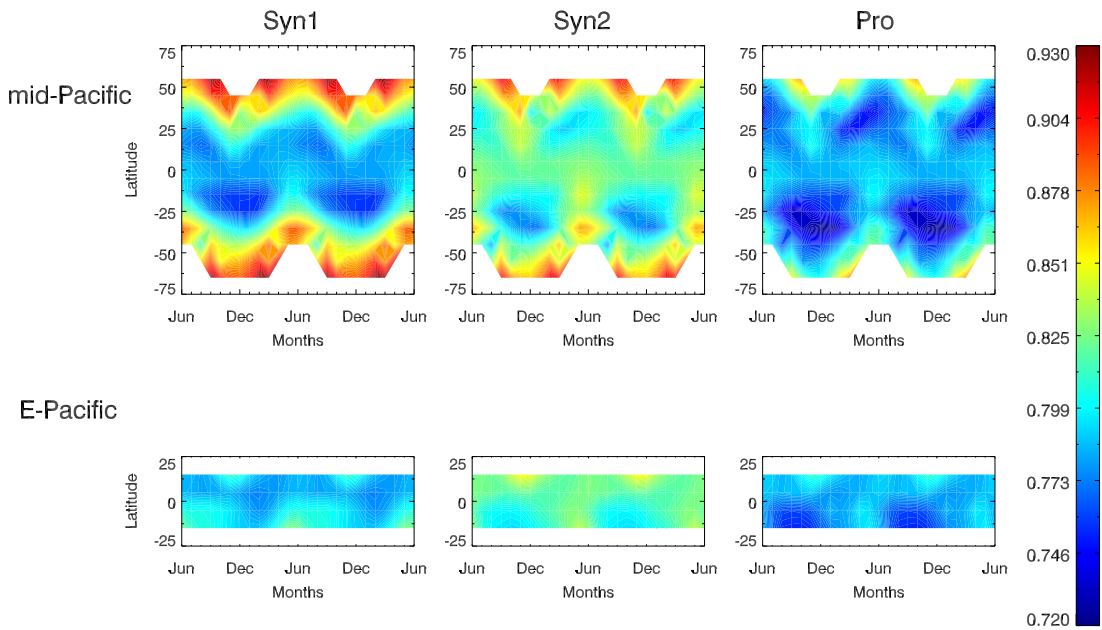

Figure 5 Multipanel time-latitude contour plots showing $P^B_Z : P^B_{Z\text{-POT}}$ ratio, mid-day depth integrated over the upper mixed layer for a 24 month period. Upper panels show zonal averages for the mid-Pacific (longitude of Hawaii) and lower panels for the east Pacific, near the equator (longitude of Galapagos, cf. Fig. 3).

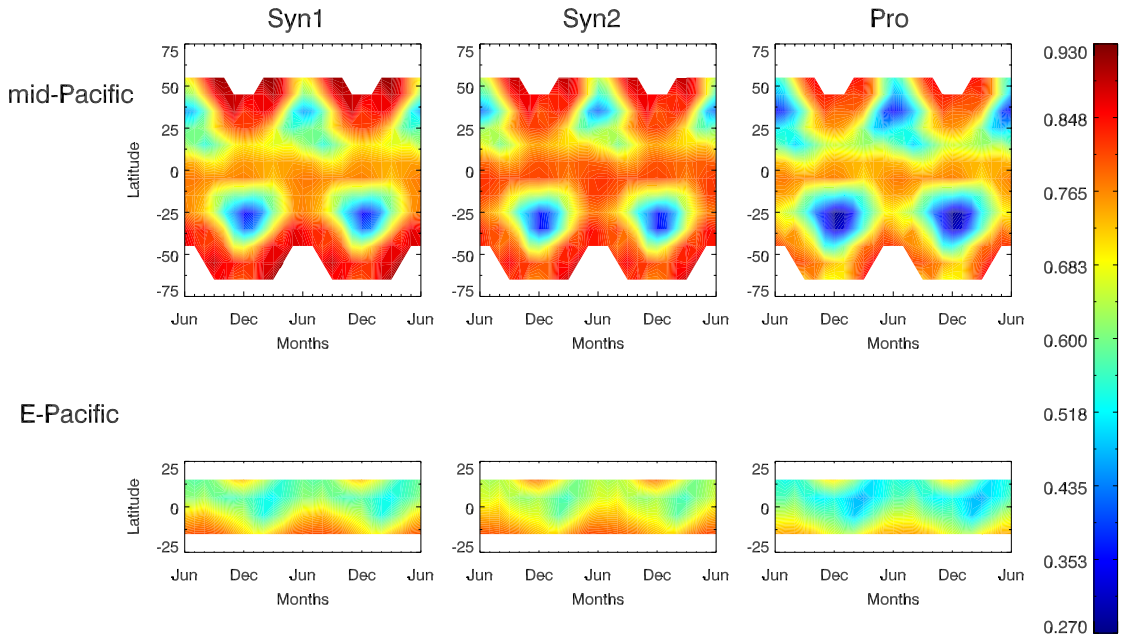

Figure 6 Effect of inhibition on weighted average productivity (monthly means, 1=January). The ratio of average hemispheric productivity including ($<P>_Z$) vs. excluding ($<P>_{\text{pot-Z}}$) inhibition due to UV and PAR over mid-Pacific and east longitudes and used to estimate average inhibition as 1-$<P>_Z/<P>_{\text{pot-Z}}$ as a percentage. Productivity at each location was weighted by cell abundance and daily PAR to be approximately proportional to each lineage's contribution to total hemispheric productivity at each time and location (see text for details). The estimates used the biological weighting functions for inhibition of photosynthesis in two strains of *Synechococcus* (Syn1 and Syn2) and *Prochlorococcus* (Pro), and were performed for productivities integrated to the 0.1% light depth for $E'_{\text{PAR}}$ (A, C) and the mixed layer depth (B, D)

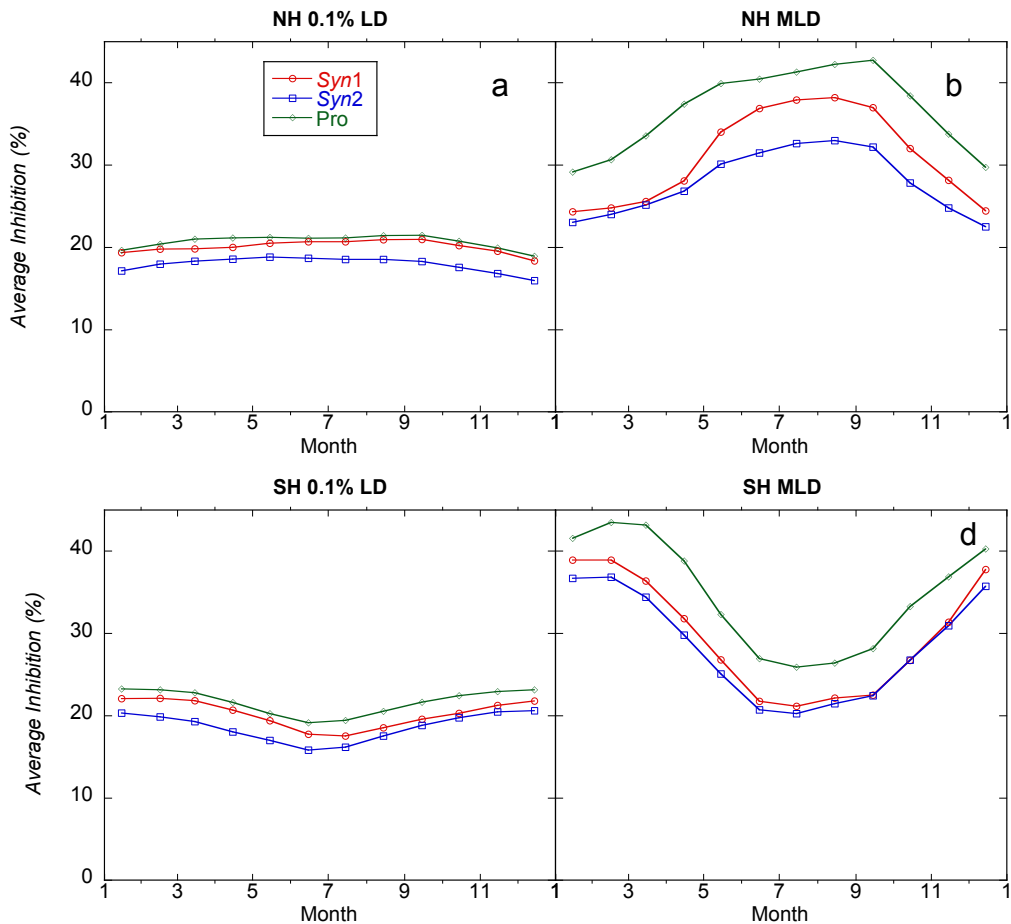

Table 1, Growth rate and photosynthetic parameters for the BWF/P-$E_{max}$ model fits for *Prochlorococcus* cultures growth under medium light (ML = 18 W m$^{-2}$/77 μmol photons m$^{-2}$ s$^{-1}$ PAR) and high light (HL = 41 W m$^{-2}$/174 μmol photons m$^{-2}$ s$^{-1}$ PAR) and temperatures of 20° or 26° C. $E^*_{max}$ is dimensionless. The last column shows exposure at which $E^*_{inh} = E^*_{max}$ as a percentage of full solar exposure, just below surface, for noon, January 17, at 15°S. Means ± SD (Growth rate) or SE (including both replicate and fit uncertainty) for n≥3 independent experiments conducted for each set of growth conditions.

| Growth Conditions | | Growth Rate d$^{-1}$ | $P^b_s$ mgC mgChl$^{-1}$ h$^{-1}$ | $E_s$ W m$^{-2}$ | $\varepsilon_{par}$ x 10$^{-3}$ (W m$^{-2}$)$^{-1}$ | $E^*_{max}$ | % of Full Sun |
|---|---|---|---|---|---|---|---|
| Light | T °C | | | | | | |
| ML | 20 | 0.28±0.03 | 1.38±0.13 | 23.26±1.34 | 1.66±0.66 | 0.85±0.12 | 13±2% |
|  | 26 | 0.33±0.01 | 2.04±0.22 | 27.94±2.52 | 1.37±0.49 | 0.98±0.11 | 17±3% |
| HL | 20 | 0.37±0.01 | 1.80±0.12 | 25.00±2.60 | 1.35±0.77 | 1.20±0.22 | 24±5% |
|  | 26 | 0.36±0.01 | 2.69±0.50 | 31.39±1.81 | 0.52±0.54 | 0.54±0.10 | 19±5% |

Supplementary Table 1, Symbols and Abbreviations

| Symbol | Name | Units |
|---|---|---|
| $a_p(\lambda)$ | Phytoplankton chlorophyll-specific spectral absorption coefficient | $m^2$ mg Chl$^{-1}$ |
| $a_{IS}(z)$ | Irradiance weighted chlorophyll-specific absorption of PAR at depth, $z$ | $m^2$ mg Chl$^{-1}$ |
| $a_{PI}$ | Irradiance weighted chlorophyll-specific absorption of PAR in the photoinhibitron | $m^2$ mg Chl$^{-1}$ |
| BWF | Biological weighting function | |
| $c$ | Scaling factor for exposures $> E^*_{max}$ | dimensionless |
| $\varepsilon(\lambda)$ | Biological weight of inhibitory effect of UV | $m^2$ mW$^{-1}$ |
| $\varepsilon_{PAR}$ | Biological weight of inhibitory effect of PAR | $m^2$ W$^{-1}$ |
| $E^*_{inh}$ | Irradiance weighted for effectiveness in inhibiting photosynthesis | dimensionless |
| $E(\lambda)$ | UV irradiance at wavelength $\lambda$ = 265-400 nm | mW m$^{-2}$ nm$^{-1}$ |
| $E^*_{max}$ | Exposure that saturates repair rate | dimensionless |
| $E_{PAR}$ | PAR irradiance (400-700 nm) | W m$^{-2}$ |
| $E_{PAR}(z)$ | Underwater PAR irradiance at depth z | W m$^{-2}$ |
| $E'_{PAR}(z)$ | Underwater PAR irradiance at depth z adjusted for difference in pigment absorption of PAR in situ vs in the photoinhibitron | W m$^{-2}$ |
| $E_s$ | Characteristic irradiance for onset of saturation of photosynthesis | W m$^{-2}$ |
| $E^Q(0^-,\lambda)$ | Spectral photon flux density of PAR (400-700 | µmol photons m$^{-2}$ s$^{-1}$ |

| | nm) at the sea surface | |
| $E_{PI}^{Q}(\lambda)$ | Spectral photon flux density of PAR (400-700 nm) in the photoinhibitron | µmol photons m$^{-2}$ s$^{-1}$ |
| $P_s^B$ | Biomass-normalized photosynthetic rate | mg C mg Chl$^{-1}$ h$^{-1}$ |
| $P_{pot}^B$ | Biomass-normalized potential photosynthetic rate in absence of inhibition | mg C mg Chl$^{-1}$ h$^{-1}$ |
| $P_s^B$ | Irradiance saturated (maximum) photosynthetic rate in absence of inhibition | mg C mg Chl$^{-1}$ h$^{-1}$ |